\documentclass[a4paper]{article}

\usepackage[margin=1in]{geometry} 

\usepackage{amsmath}
\usepackage{amsthm}
\usepackage{amssymb}

\usepackage[utf8]{inputenc}
\usepackage[hidelinks]{hyperref}
\usepackage{xspace}

\newcommand{\Ohm}{$\Omega$\xspace}
\newcommand{\half}{\frac{1}{2}}
\newcommand{\rndP}[1]{\left( #1 \right)}
\newcommand{\figref}[1]{Figure \ref{#1}}

\newcommand{\Ueff}{U_{\text{eff}}}
\newcommand{\Vth}{V_\text{th}}
\newcommand{\Eth}{E_\text{th}}
\newcommand{\Nef}{N_\text{ef}}
\newcommand{\etaRW}{\overline{\eta}_\text{v}}
\newcommand{\etaV}{\eta_\text{v}}
\newcommand{\Rsf}{R_\text{SF}}
\newcommand{\Rsfmin}{R_\text{SFmin}}
\newcommand{\EO}{\mathcal{E}_0}
\newcommand{\Ev}{\mathcal{E}_\text{v}}
\newcommand{\Es}{\mathcal{E}_s}
\newcommand{\Et}{\mathcal{E}_T}
\newcommand{\Rv}{R_\text{v}}
\newcommand{\Rs}{R_s}
\newcommand{\Cx}{C_x}

\newcommand{\etal}{et al.}

\usepackage{subcaption}
\DeclareCaptionLabelFormat{nocaption}{}

\usepackage[sort&compress,numbers,square]{natbib}

\usepackage{graphicx, color}
\graphicspath{{p/}}

\usepackage{mathrsfs} 

\usepackage{lipsum}

\title{Scaling of energy delivered through an electrostatic discharge to a small series load}
\author{\normalsize Claudia A. M. Schrama$^{1,2}$\thanks{Email address: \texttt{cschrama@mines.edu}}, Calvin Bavor$^1$, John W. Rose$^{1,3}$, P. David Flammer$^1$, Charles G. Durfee$^1$}

\date{\normalsize
	$^1$Colorado School of Mines, Physics Department, 1500 Illinois St., Golden, 80401, CO, USA\\%
    $^2$University of Pavia, Department of Physics, via Bassi 6, 27100, Pavia, Italy\\%
	$^3$Los Alamos National Laboratory, Los Alamos, 87545, NM, USA\\[2ex]%
}

\usepackage{multicol}

\begin{document}
\maketitle
	
\begin{abstract}
We study the energy delivered through a small-resistance series ``victim'' load during electrostatic discharge events in air. For gap lengths over 1~mm, the fraction of the stored energy delivered is mostly gap-length independent, with a slight decrease at larger gaps due to electrode geometry. The energy to the victim scales linearly with circuit capacitance and victim load resistance but is not strongly dependent on circuit inductance. This scaling leads to a simple approach to predicting the maximum energy that will be delivered to a series resistance for the case where the victim load resistance is lower than the spark resistance.

\vspace{5pt}
\noindent\textbf{Keywords:} Electrostatic Discharge, Energy Partitioning, Rompe-Weizel Model, Breakdown Voltage, Victim Load, Spark Resistance
\end{abstract}


\section{Introduction}
	
Advances in the design of electronic devices are often accompanied by an increased sensitivity to electrostatic discharge (ESD). The effects of ESD on a device can be direct, causing physical damage and immediate failure, or indirect, leading to delayed failure from strong or transient electric
fields~\cite{Smallwood2020,Greason1984}.
The recent popularity of wearable electronics has drawn particular attention to their ESD risk~\cite{Zhou2018}. Although voltage or current thresholds are often used as a measure of risk to electronic devices, the energy delivered by a short ESD pulse is often a good measure of risk to the device~\cite{Paasi2005}. 

In other settings, the risk extends beyond equipment damage: ESD can initiate combustion of energetic materials, resulting in serious consequences~\cite{Lovstrand1981,Larson1989,Guoxiang1982,Rizvi1992,Hearn2001,Talawar2006,Ohsawa2011,Weir2013}. The ``ESD sensitivity'', the energy required to ignite energetic material 50\% of the time, is often used as a measure of the overall sensitivity of the material. This metric has shown good agreement for different experiments with many materials~\cite{Larson1989,Skinner1998}. Still, the sensitivity to ESD can also depend on several parameters, such as chemical composition, granularity and grain shape, mechanical properties, and temperature and moisture content~\cite{Talawar2006}.

Various studies have been performed to better understand ESD events in a variety of experimental configurations, such as a spherical-planar geometry~\cite{Zaridze1996,Bach1981,Hearn2005,Mori2007,Chubb1982}. Specific attention has been given to the current delivered from charged human bodies~\cite{Katsivelis2015,Taka2009} as well as to transient phenomena, such as the effects of approach speed~\cite{Yoshida2012,Daout1987,Tomita2015} or fast voltage rise times~\cite{Parkevich2019} on peak current and peak power.

The efficiency of energy transfer in a spark discharge, defined as the amount of stored energy converted into spark energy, has been a key area of study, particularly concerning gas ignition~\cite{Zhong2015,Peng2013,Eckhoff2010,Randeberg2006}. Zhong \etal~\cite{Zhong2015} investigated the influence of different trigger methods and inductance loads (0.024~mH and 1.454~mH) on discharge efficiency and time. Their methodology involved collecting high-voltage and current probe traces to determine spark energy. They employed two primary trigger methods: a pneumatic piston that rapidly moved a grounded electrode toward a fixed high-voltage electrode, which resulted in a discharge at an unknown gap length, and a high-voltage relay that applied an overvoltage to a 6~mm gap. Within this research, only transient ESD events were studied. 

In an experimental investigation of spark discharge energy, Peng \etal~\cite{Peng2013} determined the energy delivered to the spark using an integration method of measured voltage and current in a needle-to-plate electrode configuration. Their study allowed for double current pulses, increasing the discharge efficiency to nearly 100\%. 

These studies on discharge efficiency predominantly focus on the sensitivity of explosive dust and gases to high overvoltages or transient discharge conditions. While this prior research has focused on spark-induced gas ignition under various conditions, there is a notable gap in the consistent investigation of how ESD affects the series resistive elements. This is particularly relevant for sensitive energetic materials, such as those used in explosive bridge wires (30 and 50~m\Ohm~\cite{Rae2021}), which could be exposed to an ESD from a charged tool as it is brought into proximity. In this specific scenario, the ESD event is initiated when the gap closes slowly enough that the voltage is close to the static breakdown voltage, and the gap length can be considered constant during the breakdown event; we call this a ``quasi-static'' event. Such ESD events are characterized as charge-limited at the threshold voltage. Such small resistance can also be found in experimental equipment or in more intricate circuits that make their way into the everyday household. 

In this paper, we consider a scenario in which two conducting electrodes are brought to breakdown in a quasi-static fashion, causing an ESD event in the air between these electrodes. The ``spark channel'' formed between the electrodes is treated as a resistive element in series with a ``victim'' load. The victim could be a sensitive electronic component or a combustible material. From the perspective of energy dynamics, energy is expended ionizing and exciting the gas between the electrodes, forming a plasma channel, Ohmic heating of the plasma channel, hydrodynamic expansion often forming a shock wave, and radiation of electromagnetic waves. The remaining energy is delivered to the victim load. In the classic Rompe-Weizel (RW) model, the spark resistance is assumed to change only due to the increase in electron density. All of the remaining physics is gathered into an empirical parameter ($a_R$) that corresponds to the ratio of the electron mobility to the net energy required to produce new electrons. We show that this simple model performs surprisingly well, in spite of the many mechanisms for sinking energy into the spark channel. Understanding the energy delivery to a victim load is critical for a wide range of engineering applications. This research is particularly valuable for guiding safety requirements for sensitive devices and materials with small resistance, such as those used in explosive bridge wires as well as common devices that rely on low-resistance components, such as the shunt resistors in multimeters or the circuits within inductive cooktops. The predictions based on the RW model of how energy is partitioned between the spark channel and victim load may help guide safety requirements set around sensitive devices and materials.

In the work presented here, we employ a well-characterized circuit to understand the energy transfer to a victim load under different circuit parameters (for further information see~\cite{SchramaCAM2023IoED}). Several electrode geometries and circuit capacitances, inductances, and gap lengths were tested. This allows us to gain insight into how the energy transfer scales with these parameters and to identify different physical regimes that occur for ESD events of varying stored energy. We measure current and voltage profiles over the entire discharge current pulse and use this to determine the energy delivered to the spark and victim. Building on our previous theoretical analysis~\cite{Rose2020}, we find that the spark resistance model initially put forth by Rompe and Weizel~\cite{Rompe1944} performs quite well in predicting the scaling of the energy transfer. 

\section{Material and Methods}
\begin{figure}[ht]
\centering
\includegraphics[width = 0.55\linewidth]{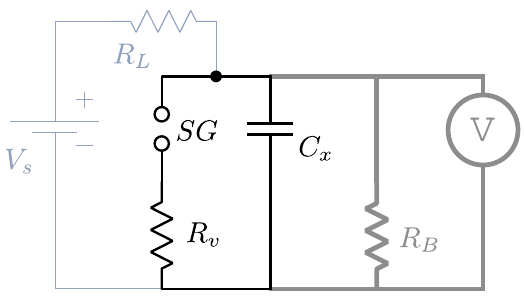}
\caption{ESD circuit diagram highlighting the charging circuit (thin, light blue), the safety circuit (thick dark gray) and the discharge circuit (black).}
\label{fig-circuit-00}
\end{figure}

The experimental setup is designed to mimic ESD events between conducting objects that approach each other slowly, so that the gap separation does not significantly change during the breakdown process. The circuit layout includes a charging circuit, safety circuit, and discharge circuit, shown in \figref{fig-circuit-00}. The external capacitor, $C_x$ (TDK low-inductance UHV series), in parallel with the spark gap branch, is charged using a switching power supply (TDK-Lambda model 500 A) through a current limiting resistor, $R_L = 100$~M\Ohm. The circuit has a hardwired high-voltage probe, $V$ (Fluke 80K-40), that indicates when the circuit is charged. The large resistance of this probe (1~G\Ohm) is in parallel with a $R_B=1$~G\Ohm resistor to dissipate any excess charge left on the circuit when power is removed. A second high-voltage probe (Cal Test Electronics CT4028) records voltage traces for the ESD events from the high-voltage side of the spark gap to ground, the dark node in the circuit diagram, see \figref{fig-circuit-00}. The ESD circuit consists of a spark gap ($SG$) in series with a low-inductance current viewing resistor (CVR), $\Rv$ (SSDN- series from T\&M Research Products). The victim load is considered to be the sum $\Rv = R_c + R_x$.

\begin{figure}[htp]
\centering
\begin{subfigure}[t]{0.3\textwidth}\subcaption{}
\centering\includegraphics[height = 2.2in]{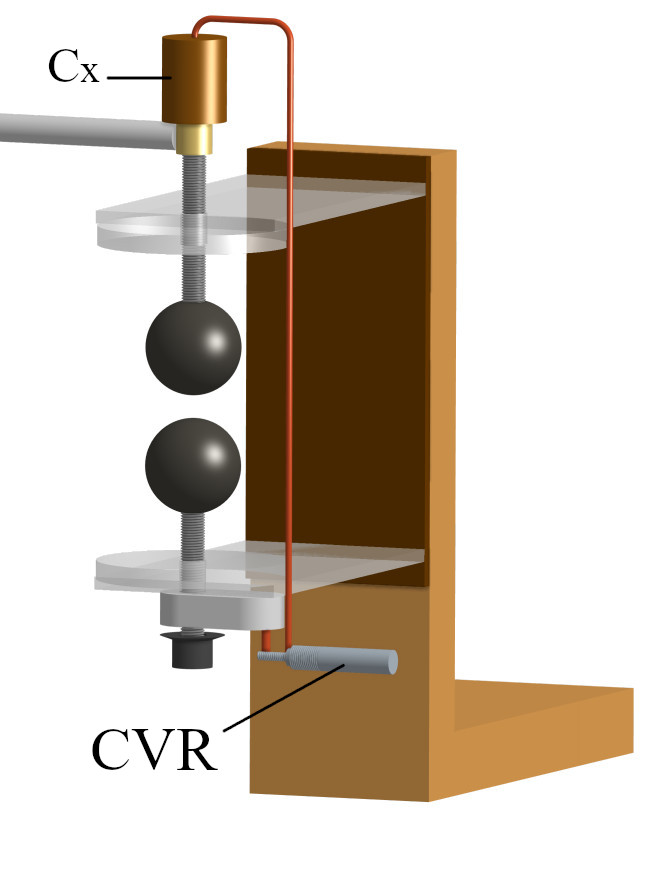}\label{fig-2a-OAS}
\end{subfigure}$\qquad$
\begin{subfigure}[t]{0.3\textwidth}\subcaption{}
\centering\includegraphics[height = 2.2in]{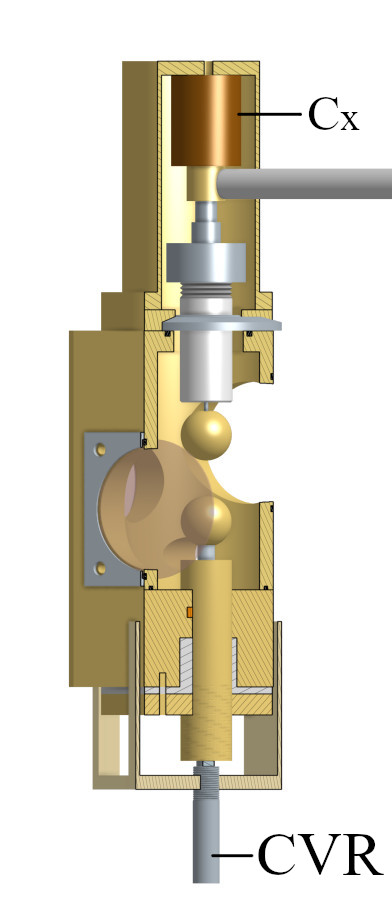}\label{fig-2b-SVC}
\end{subfigure}
\caption{Graphical depiction of the discharge circuit elements $C_x$, CVR ($\Rv$) and spark gaps of the (a) open-air system (OAS) and the (b) symmetric grounded system (SGS). For the SGS, the image is a cross-sectional view of the chamber to visualize the circuit components\label{fig-circuit-01}} 
\end{figure}

We constructed two different setups: an open-air system (OAS) and a symmetric grounded system (SGS). The SGS spark gap electrode configuration can accommodate electrodes of different shapes: 2.00~cm diameter graphite spheres, 1.27~cm diameter brass spheres, a chrome-plated steel needle (0.787~mm tip diameter), or a 1.59~cm diameter flat steel disk. The capacitances of the electrode systems were estimated to be below 10~pF. The OAS system spark gap is modified from a commercial high-voltage spark gap (Ross Engineering SG-40-H), \figref{fig-2a-OAS}. This spark gap has two 3.75~cm diameter graphite spherical electrodes separated by a variable gap length. We calculated the inductance of the OAS system to be approximately 1.1~$~\mu$H based on the frequency of the ring-down of the discharge current. This relatively large inductance results from the size of the current loop created by the return wire. Finite element simulation (using Comsol Multiphysics) of the discharge circuit gave an inductance of 0.98~$\mu$H, in reasonable agreement with the measured inductance.

The SGS system is an in-house built circuit chamber designed and manufactured for these ESD experiments. A cross-sectional view is shown in \figref{fig-2b-SVC}. The outside of the chamber was designed to provide a much more symmetric return to the ground to reduce the system inductance. During the design process, the finite element models predicted an inductance of 142~nH; the measured system inductance was 120~nH. Like the OAS, this system allows for a variable gap length and the ability to change electrodes. Since the chamber is sealed (brass and viewing windows), there is the capability to control the gas composition. For the data presented here, all gas compositions are in air at standard pressures for Golden, Colorado (roughly 630~torr or 0.83~atm at 28~$\pm$~3\% humidity). As a comparison, we obtained very similar results for a dry synthetic air ($\text{N}_2 - \text{O}_2$) mixture.

The ESD events investigated here are spontaneous, naturally initiated discharges. The circuits are charged up to the breakdown voltage, and the discharge occurs when seed electrons become present to allow the initiation of avalanche ionization. For each ESD event, we recorded the transient voltage across the spark gap - CVR branch as well as the voltage from the CVR. Voltage traces were measured with a 4-channel Tektronix MSO64 2.5~GHz oscilloscope (25 giga samples per second). In these experiments, the CVR acts as the victim load, the circuit element that receives the energy from the ESD. CVR resistances ranged from 0.00514~\Ohm to 0.25~\Ohm.

Several circuit parameters were varied to investigate the effects on energy transferred to the victim load through an ESD: external capacitance ($C_x$), electrode geometry, and inductance of the circuit (taking advantage of the different inductances of the OAS and SGS systems). On the SGS system, the external capacitance was varied from 100 to 700~pF to test how the discharge energy changes the fraction of energy delivered through the discharge channel. The SGS was also used to examine the effects of varying electrode geometry. The different geometries used were sphere-sphere, plane-sphere, and needle-plane. For asymmetric electrode geometries, the anode-cathode polarity was also switched.

\subsection{Data Processing}

\begin{figure}[ht] 
    \centering
    \begin{subfigure}{0.46\linewidth}\subcaption{}
        \includegraphics[width = \linewidth]{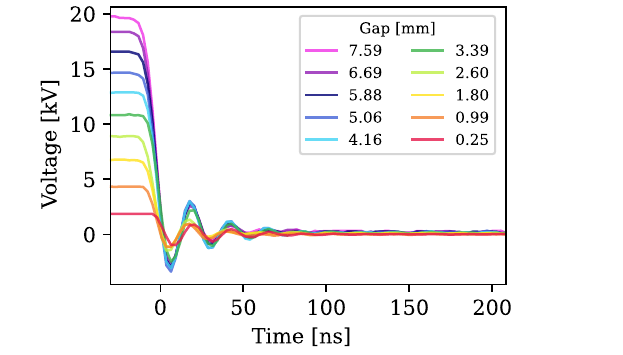} 
    \end{subfigure}
    \begin{subfigure}{0.46\linewidth}\subcaption{}
        \includegraphics[width = \linewidth]{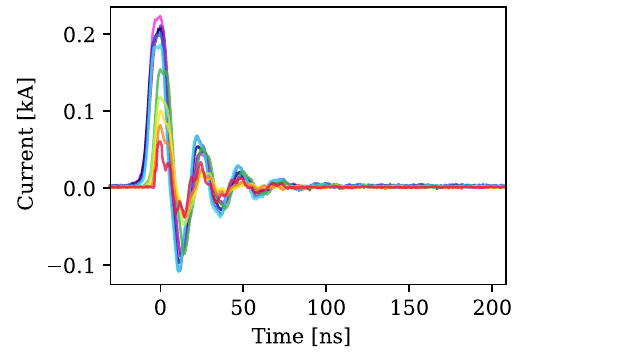}
    \end{subfigure}
    \caption{SGS example voltage and current traces with $C_x=100$~pF, $\Rv=0.0983$~\Ohm, 1.27~cm diameter brass spherical electrodes and Cal Test voltage probe. The legend in (a) applies to both plots.}
    \label{fig-02-VandITraces}
\end{figure}

Representative voltage and current traces collected using a high voltage probe (HVP) and a current viewing resistor (CVR) for various gap lengths are shown in \figref{fig-02-VandITraces}. Voltage and current data were collected for several different circuit and electrode configurations.  

\begin{figure}[b] 
    \centering
\includegraphics[width = 0.46\linewidth]{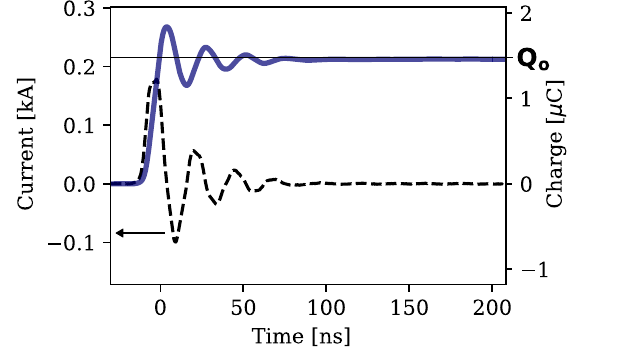} 
    \caption{The average current (dashed line) and integrated charge (solid line) for 20 individual discharge events, using the SGS with a 100~pF $C_x$ and 0.0983~\Ohm CVR at a 4.2~mm gap. The stored charge $Q_0$ is indicated by the horizontal line and marked on the charge axis.}
    \label{fig-2-Current-and-Int-Charge}
\end{figure}

Charge and energy conservation were checked on each run to validate the measurements and probe calibration. Using the measured breakdown voltage, $V_b$, and the circuit capacitance, $C$, the stored charge and energy are given by Eq.~\eqref{eq-StoredQE}; this was compared to the integrated charge and energy using the measured current, $I(t)$, given by Eq.~\eqref{eq-IntegratedQE}.
\begin{align}
    Q_0 & = C V_b & & \EO = \frac{1}{2} C V_b^2 \label{eq-StoredQE}\\
    Q_T & = \int_{t=0}^t I(t)\, \text{d}t & & \Et = \int_{t=0}^t I(t) V(t) \, \text{d}t \label{eq-IntegratedQE}
\end{align}
Here $V(t)$ is the signal from the HVP. $Q_0$ and $\EO$ are the initial charge and stored energy, respectively, corresponding to the measured breakdown voltage. $Q_T$ and $\Et$ are the charge and energy obtained by integrating the current and power delivered through the spark and victim load throughout the discharge event.

\figref{fig-2-Current-and-Int-Charge} shows the average current and integrated charge traces for 20 discharge events using the SGS with a 100~pF $C_x$ and 0.0983~\Ohm $\Rv$ at a gap length of $h=4.2$~mm. The dynamics show that the charge transfer peaks after the first half of the current cycle.

Due to differing propagation delays for the current and voltage traces, calculating the integrated $V\cdot I$ power using raw data results in an incorrect value for $\Et$ (i.e.,~not summing up to the initial stored energy). To obtain accurate integrated energies through the system that account for inductive phase shifts, it was necessary to accurately measure the cable delay difference between the respective data collection locations and the oscilloscope. These measurements were performed by measuring the transit times of fast pulses from a digital delay generator. 
  
\begin{figure}[ht] 
    \centering
    \begin{subfigure}{0.46\linewidth}
    \subcaption{}
\includegraphics[width = \linewidth]{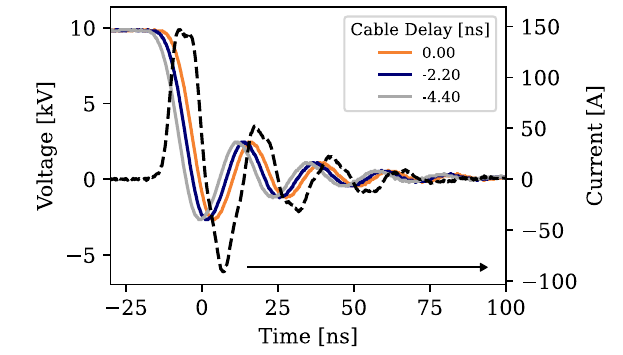} 
    \end{subfigure}
    \begin{subfigure}{0.46\linewidth}\subcaption{}
\includegraphics[width = \linewidth]{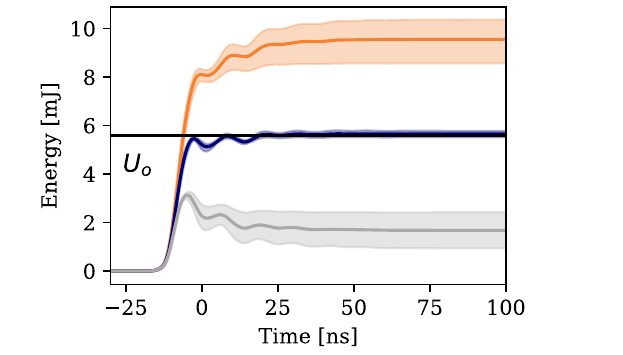}\label{fig-5b-energy}
    \end{subfigure}
    \caption{For 20 individual discharge events, shown here is the (a) voltage (solid lines) and current trace (dashed line), and the (b) dissipated energy through the circuit, using the SGS with $C_x=100$~pF and $R_v=0.0983$~\Ohm at a 3.0~mm gap. The analysis is shown for three different cable delays, emphasizing the importance of the time shift between the voltage and current signals. The shaded areas around the curves show the spread in signal, and the horizontal line in (b) indicates the stored energy. The legend in (a) applies to both plots.}
    \label{fig-2-Power-and-Int-Energy}
\end{figure}

The importance of this phase measurement is illustrated in \figref{fig-2-Power-and-Int-Energy}, which shows the power through the discharge circuit and the integrated energy for 20 discharge events using the SGS with 100~pF $C_x$ and 0.0983~\Ohm $\Rv$ at a 3.0~mm gap. \figref{fig-5b-energy} confirms that the measured phase delay of $-2.2$~ns yields the correct integrated energy. It also shows that the majority of the energy is transferred within the first half cycle of the current pulse. 

\begin{figure}[ht] 
    \centering
    \begin{subfigure}{0.46\linewidth}
    \subcaption{}
\includegraphics[width = \linewidth]{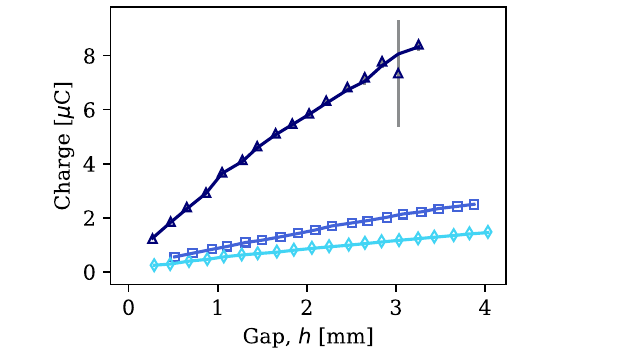} 
    \end{subfigure}
    \begin{subfigure}{0.46\linewidth}
    \subcaption{}
\includegraphics[width = \linewidth]{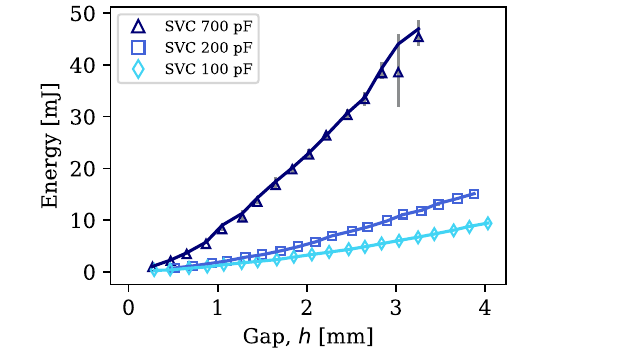} 
    \end{subfigure}
\caption{Comparison between the stored (solid line) and integrated (points) charge and energy for different external capacitors in the SGS with a 0.25~\Ohm CVR. The legend in (b) applies to both plots. There is one data point with larger error bars, which we attribute to non-ideal conditions such as dust or other dirty connections.}
    \label{fig-02-Energy-charge-conservation}
\end{figure}

The stored and dissipated charge and energy as a function of gap length are shown in \figref{fig-02-Energy-charge-conservation}, for various external capacitors in the SGS. In the figure, the stored charge and energy ($Q_0$ and $\EO$) are the solid lines, and the dissipated charge and energy ($Q_T$ and $\Et$) are the points. The conservation of charge and energy was verified for all data presented here.

For all data points presented, the reported value represents the mean of 10 to 20 unique discharge events, with the error bars indicating the standard deviation. The data point for the 3~mm discharge at 700~pF in \figref{fig-02-Energy-charge-conservation} exhibited notably larger error bars compared to the other measurements. This increased uncertainty may be attributed to a non-measurable discharge path, potentially caused by compromised electrical contacts within the circuit, such as at the electrodes or ground connections. This alternative path bypassed the CVR, which did not record the full discharge current.

With careful consideration of cable time delays and parasitic circuit inductance, the spark resistance was calculated. We found that for circuits with a small victim load resistance ($\Rv=0.0983$~\Ohm) the spark resistance reached $2.08\pm0.13$~\Ohm with $\Cx = 700$~pF, and 30--50~\Ohm with $\Cx = 100$~pF. For the case of this small victim load resistance, $\Rv \ll  \Rs$. Further details on these spark resistance measurements, including a broader range of resistance loads, are available in our related work~\cite{schrama2025variableRv,SchramaCAM2023IoED}. 

In these experiments, we used the current-viewing resistor (CVR) as our victim load. We used several different CVR resistances (0.005140 to 0.2505~\Ohm) and found that these small values did not significantly affect the discharge current. Therefore, in the following data, we normalized the energy delivered to the victim load to a victim resistance of 0.1~\Ohm. 

The victim energy, $\Ev$, is calculated using the CVR traces. The power through the victim load is $P = I_v^2\Rv$, and the total energy delivered to the victim is obtained by integrating the power, see Eq.~\eqref{eqn-Ev-Int-Power}.

\section{Theory \label{sec-theory}}

To interpret the measurements of energy transferred to the victim load, we will make use of a simple model of nonlinear spark resistance originally presented by Rompe and Weizel~\cite{Rompe1944}. The Rompe-Weizel (RW) model, along with several other circuit-based models, are frequently used as the starting point for ESD models today~\cite{Rose2020}. The RW model is based on the assumption that there is a certain energy cost per electron created by the breakdown process ($\Ueff$) and that the mobility ($\mu$) of the electrons is constant. The simplicity of models such as the RW model allows for quick calculations. The model does not explicitly consider complicated chemistry or whether energy is delivered to heat, molecular excitation, ionization, radiation, etc. The only assumption of the model is that there is an ``effective cost'' to generate an electron, which encompasses all possible energy channels. This cost is averaged over all the electrons, making the effective cost constant.

In this simple model, we assume that the electric field is uniform within the gap so that the threshold voltage for breakdown, $\Vth$, is linearly proportional to the gap length, $\Vth = \Eth h$. Here, the electric field strength for breakdown, $\Eth$, can be obtained from measured data found in Paschen curves~\cite{Paschen1889,Raizer1991}. The energy required to generate electrons in a low-temperature plasma is, in practice, much higher than the 
actual ionization energy of the background gas. For the purposes of this work, $\Ueff$ is considered an empirical parameter, as it includes a variety of energy sinks (ionization and dissociation energy, particle heating and excitation). Joule heating, $\vec{J}\cdot \vec{E}$, transfers energy from the electromagnetic field to the plasma, which is assumed to balance with the energy expended to create the electrons:
\begin{equation}
U_{\text {eff }} \frac{d n_{e}}{d t} = \vec{J} \cdot \vec{E} = \frac{J^{2}}{\sigma}=\frac{J^{2}}{e\,\mu\,n_e},
\end{equation}
where $\sigma = e\,\mu\,n_e$ is the conductivity. This equation can be integrated to find the time dependence of the electron density. We consider the plasma channel to be a uniform cylinder of length $h$ and cross-sectional area $A= \pi a^2$, with resistance, $R_S = h/(\sigma A)$, to obtain an expression for the nonlinear spark resistance. The current density, in terms of the current, is $J=I/A$. The resulting expression for the resistance is independent of $A$:
\begin{equation}
R_S(t) = 
    \rndP{
    \frac{2\, a_R}{h^2} \int_0^t I(t')^2 \mathrm{d} t'
    }^{-1/2},
\label{eq:RWresistance}
\end{equation}
where we define the Rompe-Weizel constant, $a_R \equiv e \mu/\Ueff$. Since $\Ueff$ and $\mu$ are parameters that are difficult to measure independently, $a_R$ is customarily treated as the empirical parameter of interest. In earlier work, $a_R$ has been measured in air to be between 0.5 and 2~cm$^2/$sV$^2$~\cite{Jobava2000}. 

For this research, we are interested in the energy delivered to the victim load. In the case of a uniform electric field, the initial stored energy ideally increases quadratically with $h$:
\begin{equation}
\EO= \half C \Vth^2 = \half C \Eth^2 h^2.
    \label{eq:idealE0}
\end{equation}
Departures from this simple relation will be discussed below. In the context of the RW model, the energy deposited into the spark is $\Es=\Nef\Ueff$, where $\Nef$ is the final number of electrons produced in the spark.
So, the final spark resistance, after the arc, is a function of the energy delivered to the spark:
\begin{equation}
\Rsf = \frac{h^2}{\Nef\, e\, \mu} = \frac{h^2}{\Es\, a_R}.
\end{equation}
It is convenient to define a minimum spark resistance $\Rsfmin$ for the case where all of the stored energy is used to produce electrons $(\Es = \EO)$:
\begin{equation}
\Rsfmin = \frac{2}{a_R\, C\, \Eth^2}.
\end{equation}
In this limit, the final resistance is independent of the spark gap separation, $h$. For a standard $\Eth$ of $30$~kV/cm, $C=700$~pF, and using the listed values for $a_R$, $\Rsfmin$ is calculated to be in the range of 1.59 to 6.35~\Ohm. This theoretical minimum spark resistance is therefore greater than the $\Rv$ used in these experiments.

The energy delivered to the victim load, $\Ev$, is obtained by integrating the power:
\begin{equation}
\Ev = \Rv \int_0^\infty I^2 \mathrm{d} t.\label{eqn-Ev-Int-Power}
\end{equation}
We can evaluate the nonlinear spark resistance to $t=\infty$,  Eq.~\eqref{eq:RWresistance}, to represent the integral over the current in terms of the final spark resistance to obtain
\begin{equation}
\Ev =  \Rv \frac{h^2}{2\, a_R\, \Rsf^2}.
\end{equation}
In the simple limit where the stored energy follows Eq.~\eqref{eq:idealE0}, we obtain
\begin{equation}
\Ev = \frac{\Rv}{\Rsf^2} \frac{\EO}{a_R\, C\, \Eth^2}.
\end{equation}
In the limit of small victim load resistance ($\Rv$), most of the energy is delivered to the spark, and $\Rsf = \Rsfmin$, so $\Ev = \half \frac{\Rv}{\Rsfmin} \EO$ and the fraction of stored energy delivered to the victim load is
\begin{equation}
\etaV = \frac{\Ev}{\EO} = \half \frac{\Rv}{\Rsfmin} =  \frac{1}{4}a_R\, C\, \Rv\, \Eth^2.
\label{eq:etaV}
\end{equation}
In a separate paper, we explore a more general solution for arbitrary $\Rv$~\cite{schrama2025variableRv}. Eq.~\eqref{eq:etaV} is the lowest-order term of that more general solution when expanded for small $\Rv/\Rsfmin$. When looking at the fractional energy, we can move all circuit parameters to one side to obtain
\begin{equation}
\etaRW = \frac{\etaV}{C \Rv} = \frac{a_R \Eth^2}{4}. \label{eqn-scaled-energy}
\end{equation}
This expression shows that if the fraction of stored energy is scaled by $C\,\Rv$, the result will be independent of the gap length. We will test this result in Section \ref{sec-RandD} below.  

This analysis using the RW model provides a baseline for what to expect for the energy transfer dynamics during the spark discharge. There are several important assumptions. First is making use of the relation $\Vth = \Eth h$. We find experimentally that there is an offset voltage at $h=0$ and that there are departures from linearity that arise from the geometry of the electrodes. The second assumption is that there are fixed values for the electron mobility $\mu$ and the energy cost for creating an electron $\Ueff$. We will solve for the parameter $a_R = e \mu/\Ueff$ to test this assumption. Any departures from the RW model predictions may shed light on the collisional dynamics in the plasma that are outside the bounds of the simple model. 

\section{Results and Discussion \label{sec-RandD}}
For our experiments, we measured the current and voltage profiles for different values of the capacitance, inductance, electrode geometry, and gap length.  

\subsection{Gap length dependence of breakdown voltage}

\begin{figure}[ht] 
    \centering
    \begin{subfigure}{0.46\linewidth}
    \subcaption{}
        \includegraphics[width = \linewidth]{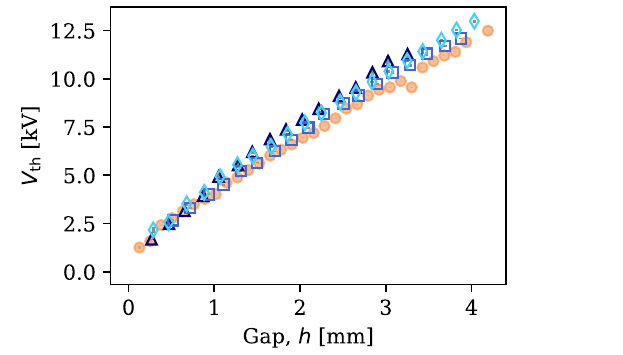} \label{fig-R_and_D-Cap_and_Inductance-a}
    \end{subfigure} 
    \begin{subfigure}{0.46\linewidth}
    \subcaption{}
\includegraphics[width = \linewidth]{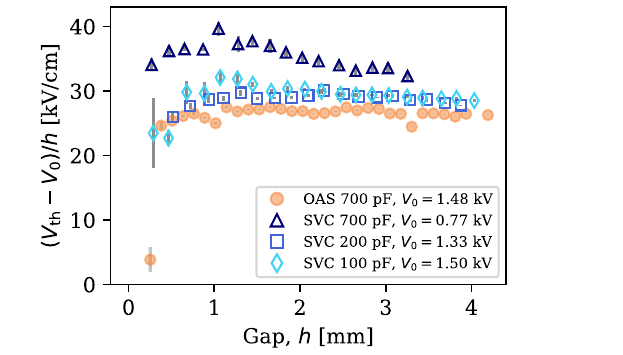} \label{fig-R_and_D-Cap_and_Inductance-b}
\end{subfigure}
\caption{Demonstrating the effect of varying capacitance on (a) breakdown voltage, and (b) breakdown field strength for the SGS with $C_x$ $=$ 100, 200, and 700~pF, $\Rv= 0.25$~\Ohm, and brass 1.27~cm diameter electrodes. The OAS data ($C_x$ $=$ 700~pF, $\Rv= 0.00514$~\Ohm, 3.75~cm graphite sphere electrodes) is overlaid for comparison with the SGS 700~pF trace to observe the effects of circuit inductance. All data points are averaged over 10 discharge events per gap length and normalized to the victim load size of 0.1~\Ohm. The legend in (b) applies to both plots.}
\label{fig-CAP-IND-Vb_Eth}
\end{figure}

Since the primary objective of this project is to understand what fraction of the stored energy is delivered to a victim load, we made measurements of the breakdown voltage as a function of the electrode gap for all of the discharge conditions. \figref{fig-R_and_D-Cap_and_Inductance-a} shows the measured voltage just before the start of the discharge as a function of gap separation for the SGS system that had external capacitance values of $C_x=100$, 200 and 700~pF (shown in shades of blue) and 1.27~cm diameter brass spherical electrodes, as well as the OAS system where the capacitance was 700~pF (orange circles) and 3.75~cm diameter graphite spherical electrodes. The trends are close to linear in $h$ as expected: \figref{fig-R_and_D-Cap_and_Inductance-b} shows the breakdown field calculated by fitting the $h < 2~{\rm mm}$ section of the data to a line. In fitting the data, we find voltage offsets of approximately $V_0=1.45$~kV at $h=0$. Similar offsets have previously been observed in the literature~\cite{Hogg2015,Meng2019} and likely result from a dielectric layer on the electrodes, such as an oxide layer~\cite{Farber2023}. The breakdown fields range from 28--34~kV/cm. The breakdown voltage is expected to vary with electrode geometry and material but can also depend on ambient air pressure and humidity, which was not controlled in our experiment. In principle, all of the data for the SGS chamber in \figref{fig-R_and_D-Cap_and_Inductance-a}should show the same $V_th$; the slight observed variability could result from conditioning of the electrode surfaces during the experiments. The data for the OAS (graphite electrodes) show a slightly lower breakdown threshold than for the SGS (brass electrodes) which we attribute to the electrode material.

\begin{figure}[ht] 
\centering 
\begin{subfigure}{0.46\linewidth}
    \subcaption{}
    \includegraphics[width = \linewidth]{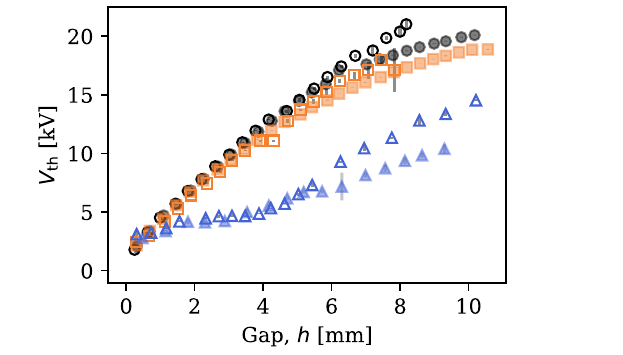} \label{fig-GEOMETRY-Vbreak}
\end{subfigure} 
\begin{subfigure}{0.46\linewidth}
    \subcaption{}
    \includegraphics[width = \linewidth]{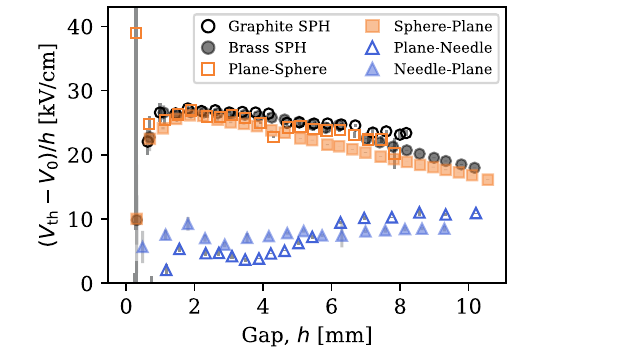} \label{fig-GEOMETRY-Eth}
\end{subfigure}
\caption{Demonstrating the effect of varying electrode geometries on (a) breakdown voltage, and (b) breakdown field strength for the SGS with $C_x= 100$~pF and $\Rv = 0.1$~\Ohm. All data points are averaged over 10 discharge events per gap length. Similar $V_0$ offsets were found with a fit on the $\Vth$ data with a range between 1.3 and 1.8~kV. The legend in (b) detailing anode-cathode electrode pairings applies to both plots.}
\label{fig-GEOMETRY-Vb-Eth}
\end{figure}

In \figref{fig-GEOMETRY-Vb-Eth}, we show similar data for the breakdown voltage vs. gap length for various electrode geometries over a broader range of gap lengths. The SGS system was used with $C_x=100$~pF and 0.10~\Ohm CVR. The legend for the different electrode geometries follows the notation anode-cathode; for example, ``needle-plane'' indicates a needle-shaped anode and a planar cathode.

For the non-needle electrodes, there is a linear increase in breakdown voltage of approximately 26~kV/cm at smaller gap lengths. The breakdown voltage is lower than the linear trend for larger gap distances. We interpret this as due to the fact that for large gap lengths, the field is less uniform and concentrated near the electrodes. The needle-shaped electrode, either as the anode or cathode, emphasizes this effect, significantly reducing the breakdown voltage; this effect reduces the necessary potentials to initiate discharge events in gaps with strongly non-uniform electric field distributions~\cite{Bazelyan1998}.

\subsection{Varying circuit capacitance}

For a given gap separation, an increase in capacitance increases the initial stored energy and charge. The energy transferred to the victim load $\Ev$ is shown in \figref{fig-CAP_IND-E_v}. As expected, $\Ev$ increases with $h$ and $\Cx$. As detailed in Section \ref{sec-theory}, much of the variation shown can be accounted for in the parameter $\etaRW$ that normalizes $\Ev$ to the $\EO$, $\Rv$ and $\Cx$, see Eq.~\eqref{eqn-scaled-energy}. \figref{fig-CAP_IND-EtaRW}, shows that for $h>1~{\rm mm}$, $\etaRW\approx 0.4~\Omega^{-1}$nF$^{-1}$ for all of the tested capacitances. The fraction of stored energy transferred to the victim load is quite small: even for $\Cx = 700$~pF, $\etaV$ is only 2.8\%. This confirms the utility of the Rompe-Weizel model. There is, however, a rise in $\etaRW$ at lower gaps for the lower stored energy cases. This deviation from a constant results from the voltage offset at $h=0$ giving a stored energy that is larger than expected at small gaps.

The rise in $\etaRW$ with smaller gaps seen in \figref{fig-CAP_IND-EtaRW} appears to be related to the observed breakdown voltage offset, $V_0$: the stored energy has a component $\EO=\frac{1}{2}C V_0^2$ that is independent of $h$. When we account for this voltage offset, we can use a different approach to plotting the data. Eq.~\eqref{eq:etaV} can be adjusted to replace $\Eth$ with $(V_0+\Eth h)/h=\Vth/h$, where $\Vth$ is the observed breakdown threshold voltage. Making use of this relation, we can instead plot a calculated $a_R(h)$:
\begin{equation}
a_{R}=\frac{4\, \etaV\, h^{2}}{\Rv C_{x} \Vth^{2}}.
    \label{eq:aRvsh}
\end{equation}
\figref{fig-CAP_IND-aR} shows this calculated value of $a_R(h)$. The values are fairly constant at larger gap lengths, indicating that the voltage offset is likely responsible for the small-gap divergence in \figref{fig-CAP_IND-EtaRW} for the lower capacitance. Using the SGS, we measured discharges as a function of pressure of synthetic dry air, finding that the value of $a_R$ is inversely proportional to the pressure, which is consistent with the expected density dependence of the electron mobility. A variation of air pressure of ${\pm}4\%$ corresponds to a variation of the value of $a_R$ by ${\pm}0.075$, small compared to the variation seen in \figref{fig-CAP_IND-EtaRW}. The slight decrease at smaller gaps for the larger capacitances indicates that for these cases, the larger discharge current may affect the condition of the electrode surfaces to reduce the voltage offset. The $a_R$ values are in a range similar to those observed by Jobava \etal~\cite{Jobava2000} that are indicated in the shaded region. 

\subsection{Varying circuit inductance}

\begin{figure}[htp]
\captionsetup[subfigure]{labelformat=nocaption}
\centering
\includegraphics[width=0.95\linewidth]{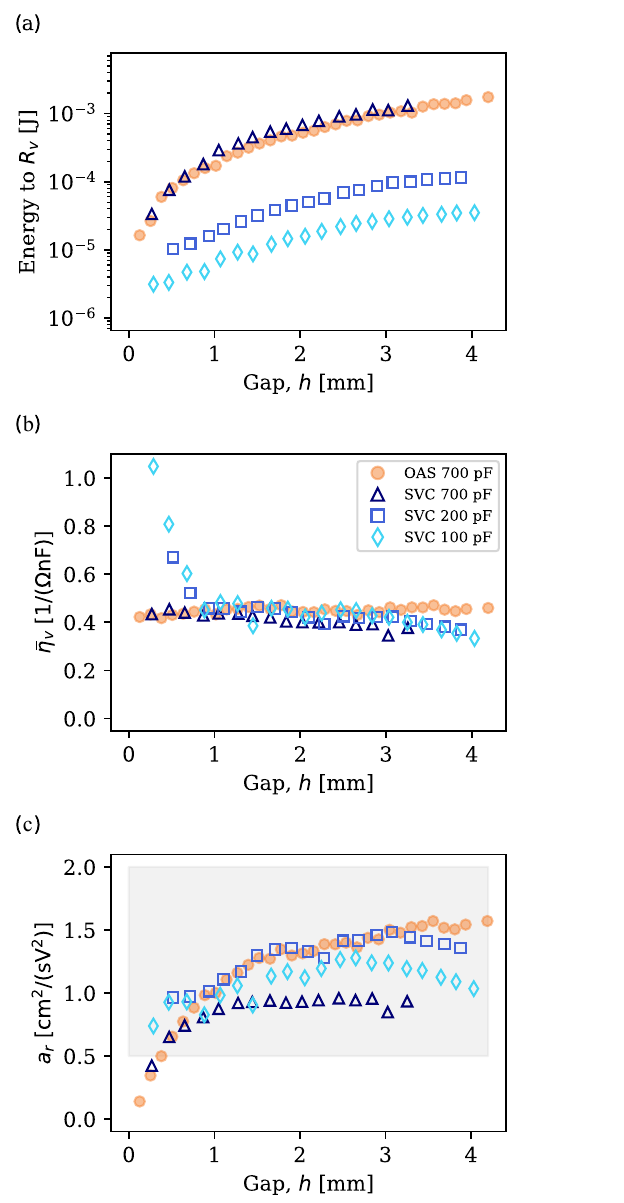}
\begin{subfigure}{0\linewidth}\caption{}\label{fig-CAP_IND-E_v}\end{subfigure}
\begin{subfigure}{0\linewidth}\caption{}\label{fig-CAP_IND-EtaRW}\end{subfigure}
\begin{subfigure}{0\linewidth}\caption{}\label{fig-CAP_IND-aR}\end{subfigure}
\caption{Demonstrating the effect of varying capacitance on (a) energy dissipated by $\Rv$, (b) $\etaRW$, and (c) $a_R$ for the SGS with $C_x$ $=$ 100, 200, and 700~pF, $\Rv=0.25$~\Ohm, and 1.27~cm diameter brass sphere electrodes. The OAS data ($C_x$ $=$ 700~pF, CVR $=$ 0.00514~\Ohm, 3.75 graphite sphere electrodes) is overlaid for comparison with the SGS 700~pF trace to observe the effects of circuit inductance. All data points are averaged over 10 discharge events per gap length and normalized to the victim load size of 0.1~\Ohm. The legend in (b) applies to all three plots.}
\label{fig-GEOMETRY-Ev_EtaRW_aR}
\end{figure}

The influence of circuit inductance on discharge dynamics was investigated by operating the SGS (124~nH, 1.27~cm diameter brass spherical electrodes) and OAS (1.1~$\mu$H, 3.75~cm diameter graphite spherical electrodes) systems with the same 700~pF external capacitor. The data for the OAS system is shown in orange circles in \figref{fig-CAP-IND-Vb_Eth}. The breakdown voltage is seen to be similar, with a slightly higher offset voltage, likely resulting from the different electrode materials. Furthermore, the energy dissipated by the victim load was comparable between the SGS and OAS configurations, \figref{fig-CAP_IND-E_v}. 

\figref{fig-CAP_IND-EtaRW} displays the normalized energy transfer to the external capacitor and victim load, calculated using Eq.~\eqref{eqn-scaled-energy}. For both the 700~pF SGS and OAS datasets, the normalized energy remained relatively constant across the gap lengths tested. This agreement in discharge dynamics between the two systems indicates that a tenfold increase in inductance does not significantly impact the energy delivered to the victim load. Since most of the stored energy is dissipated into the spark in the early phase of the discharge, the spark resistance during the latter phase of the discharge is constant. In this case, where the dynamics are in a linear regime, $\etaV$ would be expected to be independent of the inductance.

\twocolumn

\begin{figure}[htp]
\captionsetup[subfigure]{labelformat=nocaption}
\centering
\includegraphics[width=0.95\linewidth]{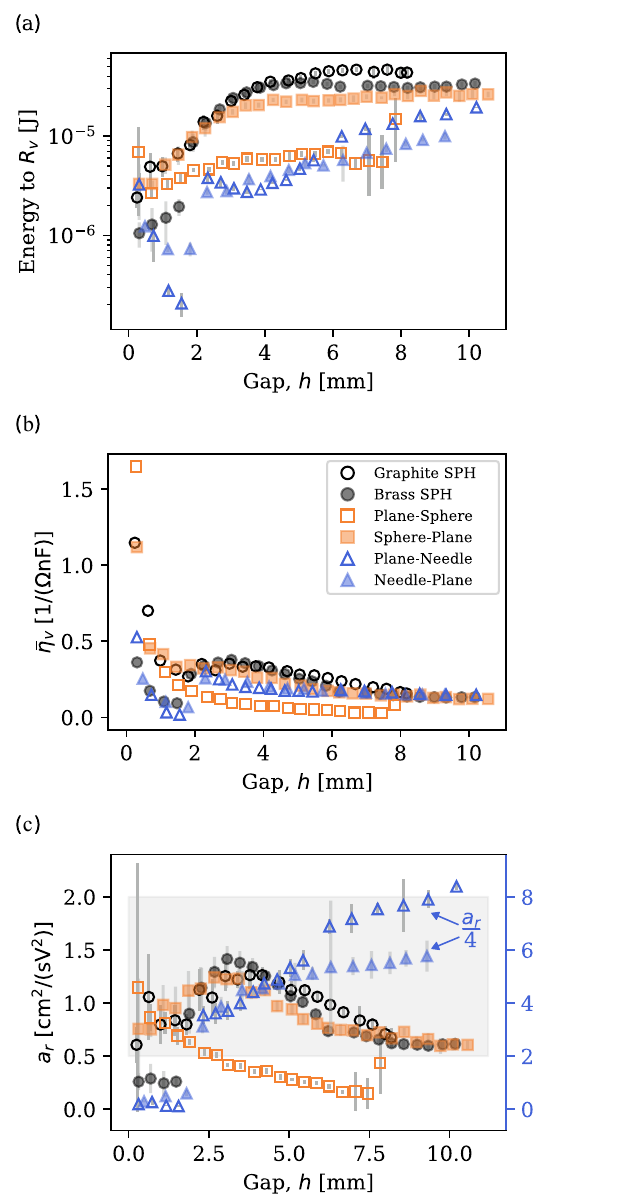}
\begin{subfigure}{0\linewidth}\caption{}\label{fig-R_and_D-Geometry-a}\end{subfigure}
\begin{subfigure}{0\linewidth}\caption{}\label{fig-R_and_D-Geometry-b}\end{subfigure}
\begin{subfigure}{0\linewidth}\caption{}\label{fig-R_and_D-Geometry-c}\end{subfigure}
\caption{
    (a) Energy dissipated by $\Rv$, (b) $\etaRW$, and (c) $a_R$ for the SGS with $C_x = 100$~pF and $\Rv=0.1$~\Ohm, demonstrating the effect of varying electrode geometries. All data points are averaged over 10 discharge events per gap length. The legend in (b) detailing anode-cathode electrode pairings applies to all plots. In (c), the data in blue triangles involving a needle electrode have been divided by a factor of 4; their true scale is labeled on the right side of the plot.}
\label{fig-R_and_D-Geometry}

~

~
\end{figure}

\onecolumn

\subsection{Varying electrode geometry}

The spherical-electrode geometry for small gaps leads to relatively uniform fields between the electrodes. When we increase the gap length and vary the electrode geometry, the shape and finite size of the electrodes result in non-uniformity of the field, as we saw in the breakdown voltages in \figref{fig-GEOMETRY-Vb-Eth}. The consequence of these effects can be seen in \figref{fig-R_and_D-Geometry-a}, which shows the energy delivered to the victim load as a function of gap length for several combinations of electrodes and polarities. The electrodes featuring a needle-shaped electrode and the orientation plane-sphere have lower absolute energy transfer and peak current to the victim. In the plane-sphere geometry, we see lower energy transfer when the cathode is spherical: while this difference is not currently fully understood, it should be noted that in this case the electrodes are made of different materials.  

\figref{fig-R_and_D-Geometry-b} is the counterpart to \figref{fig-CAP_IND-EtaRW}, showing how the fraction of stored energy transferred to the victim load scaled by $C\,\Rv$ varies with gap length. The figure clearly shows that even here, the $\etaRW$ scaling results in a reasonably constant value for $h>2~{\rm mm}$, in spite of the different discharge dynamics. Interestingly, while the RW model predicts $\etaRW$ to be a constant with no gap length dependence, each geometry shows an initial steep decrease for gaps below 1~mm, followed by a more gradually down-sloping linear relationship at longer gap lengths. The variability in the curve, which differs from a straight line, indicates that more interesting physics is present than the RW model represents, especially at the smaller gap lengths.

Plotting $a_R(h)$, \figref{fig-R_and_D-Geometry-c}, using Eq.~\eqref{eq:aRvsh}, shows values similar to those shown earlier for the geometries with a spherical electrode, but the calculated values of $a_R$ are somewhat larger with the needle geometry (in the figure these values have been divided by 4). Since $a_R = e \mu/\Ueff$, a larger $a_R$ would result from a larger mobility or smaller energy cost for producing an electron. 

\section{Conclusions}

In this work, we considered quasi-static discharge events in air (the voltage was raised in a quasi-static fashion while the gap length was held constant), with a series victim load of relatively small resistance (0.1--0.25~\Ohm). By measuring both current and voltage profiles as functions of time, we characterized the fraction of stored energy being delivered to the victim load. These measurements were made with several capacitances and gap lengths (which varied the stored energy), two different values of inductance, and several electrode geometries. We test the classic Rompe-Weizel model with our experimental results, finding that the simple model works better than one would think, considering all of the mechanisms that contribute to the empirically-measured $a_R$ parameter. The range of $a_R$ values we find are consistent with what has been seen in previous work. The variations of the $a_R$ parameter shows that deeper investigation is warranted. In ongoing work, we are developing models that more explicitly account for the underlying physics.

With the spherical electrode geometry, varying the capacitance, gap length, and inductance for gap lengths larger than 1~mm produced remarkably consistent results: the fraction of stored energy delivered to the victim load was approximately proportional to both the load resistance and the storage capacitance, and was independent of the gap length (\figref{fig-CAP-IND-Vb_Eth}). The two spherical electrode types tested were of different radius and material; we did not see a strong dependence of our results connected to the sphere radius, indicating that the independence of the results with gap length is be due to the sphere radius being sufficiently large compared to the gap length. The simple theory Rompe-Weizel model led to $a_R$ values between 0.8--1.5~cm$^2$/sV$^2$, consistent with what has been reported elsewhere in the literature.

At low gap lengths, a significant deviation from the established scaling of $\etaRW$ was observed, as the ratio of stored energy delivered to the victim load was greater than anticipated. For the smaller capacitances (100--200~pF), this can be explained by an offset in the breakdown voltage at a zero-gap length (see \figref{fig-R_and_D-Cap_and_Inductance-a}), which indicates that the stored energy does not approach zero as the gap length is reduced. We hypothesize that this breakdown voltage offset is caused by the presence of thin dielectric layers and/or surface texture on the electrodes. For the larger capacitance (700~pF), the behavior for $\etaRW$ follows the simpler model well. Our conjecture is that repeated discharges at the larger currents associated with the increased stored energy could reduce or remove this oxide layer. The role played by electrode surface conditions on the breakdown voltage merits further investigation, which is beyond the scope of this current paper. Nevertheless, it is worth noting that the deviations from the expected behavior occur under conditions where the initial stored energy (and therefore the energy transferred to the victim load) is the lowest.

Varying the geometry produces more deviation from simple theory. In particular, the sharp features in the needle-plane/plane-needle geometries significantly decrease the breakdown voltage and energy delivered to the victim. The field concentration at the sharp electrode allows the field to reach breakdown levels at lower potentials. We find that the $a_R$ parameter is substantially higher for this geometry. Preliminary imaging studies in our lab~\cite{SchramaCAM2023IoED} indicate that the channel radius is likely smaller for these ESD events. The deposition of energy into a smaller channel volume likely leads to significantly higher mobility or lower energy cost to generate electrons.

Our results here provide simple scaling for energy delivery: the fraction of the stored energy delivered to the victim is $\etaV < 0.5\ (\Omega \mathrm{nF})^{-1} {C \Rv}$, where $C$ is the capacitance of the system, and $\Rv$ is the resistance of the victim. As an example, on the high end of the parameters studied here, a 700~pF capacitance and a 0.25~\Ohm resistance result in approximately 8\% of the total stored energy to be delivered to the victim load, independent of gap lengths studied. While the risk that a given absolute energy that is delivered to a sensitive component must be assessed in each situation, we can say that when the resistance of that component is small compared to the resistance of the spark, the vast majority of the stored energy will be deposited into the spark rather than the victim load. Even so, increases in the capacitance and gap distance will generally lead to larger amounts of stored energy, resulting in an increase in the absolute energy delivered to the victim load and, therefore, an increase in the risk to components. In this work, we present investigations of energy transfer in the limit where the victim load resistance is small. In a related paper~\cite{schrama2025variableRv}, we have investigated the scenarios for arbitrary victim load resistances, where the energy partition can skew toward much higher fractions of stored energy delivered to the victim load. For the small victim loads studied here, the upper bound of energy delivery can provide guidance for energy risk tolerances.


\section{Acknowledgments}
We gratefully acknowledge financial support for this work from Los Alamos National Laboratory. We also acknowledge very useful conversations with Jonathan Mace, Dan Borovina, Travis Peer, and Francis Martinez at Los Alamos National Laboratory. We would also like to thank Forrest Doherty for his modeling of the inductance of our discharge systems during his time at Mines. The work in this paper builds on the PhD thesis by C. Schrama~\cite{SchramaCAM2023IoED}.

\bibliographystyle{ieeetr} 
\bibliography{biblio}

\end{document}